\documentclass[twocolumn,12pt]{article}
\usepackage[margin=.9in]{geometry}
\usepackage{graphicx}
\usepackage{mathtools}
\usepackage{esint}
\usepackage{amsmath, amsthm, amsfonts, amssymb}
\usepackage{todonotes}
\usepackage{hyperref}
\usepackage{nicefrac}
\usepackage{subcaption}
\usepackage{booktabs} 
\usepackage{fancyvrb} 
\usepackage{gensymb, textcomp} 
\usepackage{dblfloatfix} 

\usepackage[utf8]{inputenc}
\usepackage[T1]{fontenc}
\usepackage{blindtext}

\usepackage{float} 
\usepackage{dblfloatfix} 

\DeclareCaptionLabelFormat{adja-page}{\hrulefill\\#1 #2 \emph{(previous page)}} 

\title{Climate Modelling in Low-Precision: 
Effects of both Deterministic \& Stochastic Rounding.
}
\author{
E.\ Adam Paxton, 
Matthew Chantry, 
Milan Kl\"ower,
Leo Saffin
\& Tim Palmer
}

\graphicspath{{images/}}

\newcommand{\RR}{\mathbb{R}}

\newcommand{\EE}{\mathbb{E}}
\newcommand{\WW}{\textrm{WD}}
\newcommand{\AAA}{\mathcal{A}}
\newcommand{\dd}{\partial}

\newcommand{\eq}[1]{\begin{align} #1 \end{align}}
\newcommand{\eqq}[1]{\begin{align}\begin{split}#1\end{split}\end{align}}
\newcommand{\al}[1]{\begin{align*} #1 \end{align*}}

\begin{document}


\twocolumn[
	\begin{@twocolumnfalse}
		\maketitle
	\begin{abstract}
		Motivated by recent advances in operational weather forecasting, 
we study the efficacy of low-precision arithmetic for climate simulations. 
We develop a framework to measure rounding error in a 
climate model which provides a stress-test for a low-precision 
version of the model, and we apply our method to a variety of models 
including the Lorenz system; a shallow water approximation for turbulent flow 
over a ridge; and a coarse resolution global atmospheric 
model with simplified parameterisations (SPEEDY). 
Although double precision (52 significant bits) is standard across 
operational climate models, in our experiments we find that 
single precision (23\,sbits) is more than enough and that 
as low as half-precision (10\,sbits) is often sufficient. 
For example, SPEEDY can be run with 12\,sbits across the entire 
code with negligible rounding error and this can be lowered to 10\,sbits
if very minor errors are accepted, amounting to less than 0.1\,mm/6hr for the 
average grid-point precipitation, for example. 
Our test is based on the Wasserstein 
metric and this provides stringent non-parametric bounds on rounding 
error accounting for annual means as well as extreme weather 
events.
In addition, by testing models using both round-to-nearest (RN) and 
stochastic rounding (SR) we find that SR can 
mitigate rounding error in both
turbulent regimes and slow processes such as heat diffusion in 
a land-surface component, where low-precision is notably problematic 
with RN due to stagnation. 
Thus our results also provide evidence that SR could be relevant to 
next-generation climate models.
While many studies have shown that low-precision arithmetic can be suitable on
short-term weather forecasting timescales, our results give the first evidence that
a similar low precision level can be suitable for climate. 
I.e. In our applications, we find that small rounding errors in the short-term 
forecast do not accumulate in time so as to 
disrupt the long-time statistics of the system.
		\vspace{.25cm}
	\end{abstract}
\end{@twocolumnfalse}
]


\section{Introduction}


Modern numerical earth system models
require enormous amounts of computational resources
and place significant demand on the world's most powerful supercomputers.  
As such, operational forecasting centres are stretched to make best use 
of resources and seek ways of reducing unnecessary computation and 
memory allocation for the sake of performance wherever possible. 

One idea to improve computational efficiency which has gained attention in 
recent years is to utilize low-precision arithmetic---in place of conventional 64\,bit 
arithmetic---for computationally intensive parts of the code. 
This has been accompanied by parallel trends in deep learning
where low precision is deployed routinely and for which novel hardware is now
emerging \cite{Gupta2015, micikevicius2018mixed}.
Whether such hardware can be exploited for weather \& climate, 
however, ultimately depends on the cumulative effect of rounding error.
In fact, a number of studies have shown that much numerical weather prediction, 
at least on the short timescales relevant for forecasts, can be 
optimized for low precision \cite{Chantry2019, Hatfield2019, Jeffress2017, Kloewer2020JAMES, Saffin2020} 
and forecasting centres are already exploiting this in operations. 
The European Centre for Medium-Range Weather Forecasts 
has now ported the atmospheric component of its 
flagship Integrated Forecast System to single precision \cite{ECMWFsingle, Vana2017}
while MeteoSwiss and the UK Met Office 
have tested single and mixed-precision codes respectively \cite{Gillham2018, cosmo2014}.

As operational weather forcasters experiment with more efficient low-precision hardware, it is 
natural to ask whether low precision is suitable for climate 
modelling (i.e. long time-scales) and  
this is the question addressed by the current paper. 
Compared to weather forcasting, where research in low precision
has focused to date, climate modelling 
presents a different problem requiring some new techniques. 
While an ensemble weather forecast seeks a relatively localized
probability distribution over
the possible states of the atmosphere at a given time, 
the exact state is understood to be totally unpredictable on long 
timescales due to chaos, 
and a climate model
seeks instead to approximate the statistics of states 
over a long time period
(in the language of ergodic theory, the climatological 
object of interest is the invariant probability distribution). 
Thus the test for a low-precision climate model should be 
whether it has the 
same statistics (invariant distribution) as its high-precision counterpart. 


\begin{figure}
\centerline{\includegraphics[width = 0.9\linewidth]{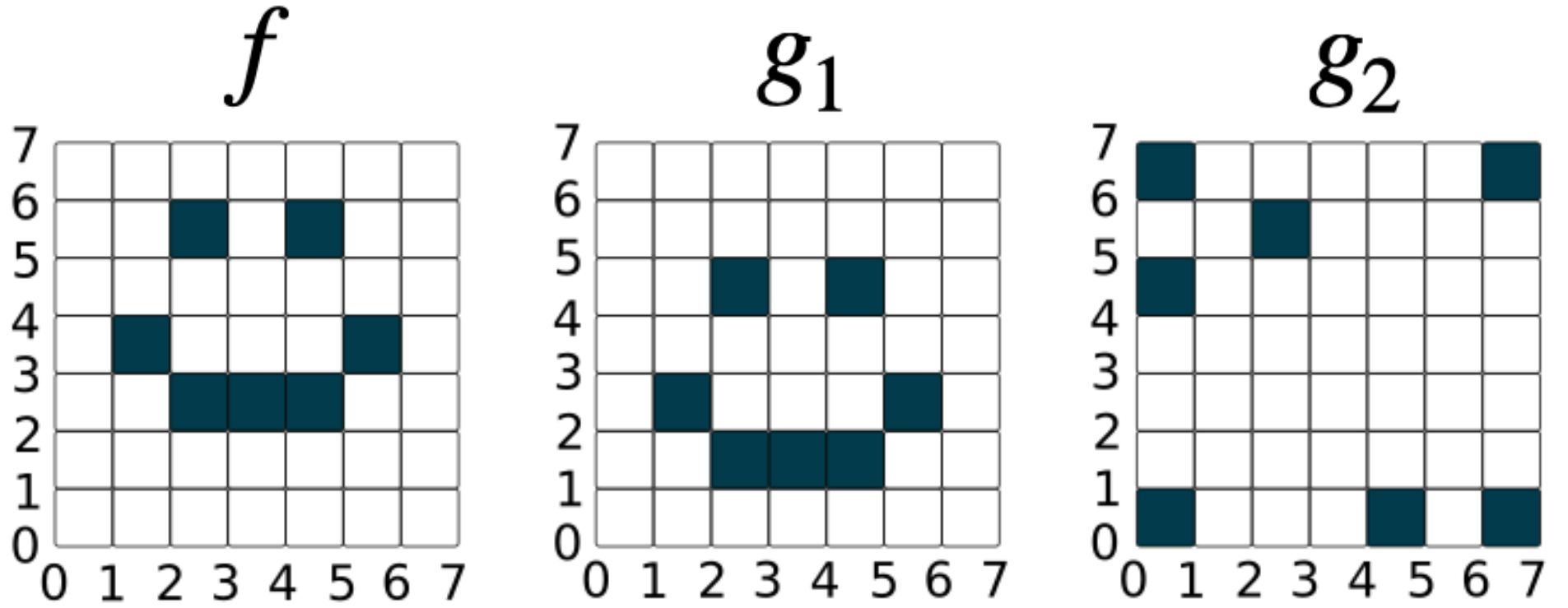}}
\captionsetup{width=.9\linewidth}
\caption[width=.95\linewidth]{An example of the Wasserstein distance (WD). 
The WD from $g_1$ to $f$ is 1, 
where an optimal transport strategy is to shift all of the probability 
mass up by one unit. 
By contrast, it is easily seen that $\text{WD}(f,g_2)>1$, reflecting the intuition 
that $g_1$ is a better approximation of $f$ than is $g_2$. 
Figure motivated by \cite{Robin2017}.}
\label{fig:smiley}
\end{figure}


We develop such a test based on the Wasserstein distance (WD) from 
optimal transport theory 
which provides a natural notion of closeness between probability distributions. 
The WD is defined as the cost of an optimal strategy for transporting 
probability mass between two distributions with respect to a cost $c(x,y)$
of transporting unit mass from $x\in \RR^d$ to $y\in \RR^d$, 
where throughout this 
paper we take $c(x,y)=|x-y|$ so that cost has the same units as the 
underlying field. 

The WD is an appropriate metric for this study because:
(1) it is non-parametric; 
(2) it has favourable geometrical properties (cf.\ Fig.\ \ref{fig:smiley}); 
(3) it is interpretable in appropriate physical units;
and (4) it bounds a range of expected values 
covering both the mean response and 
extreme weather events. 
The metric is popular in machine learning \cite{WGAN17} and 
has recently been suggested as an appropriate measure of skill
in climate modelling \cite{Robin2017, Vissio2020, Vissio2018},
however since it is not so well-known within the community, 
a survey of the WD
including a rigorous definition and discussion of computational techniques
are given in App.\ \ref{app:Wass}.
In particular, the reader is encouraged to see \ref{WDproperties} for 
further discussion of (1)-(4).

Any 
test can only bound the effects of rounding error at low precision 
relative to the variability of probability distributions generated by
a corresponding high-precision experiment. 
Experiments must thus be carefully designed to minimise such variability in 
order to isolate the effects of rounding error. 
For example, by taking an ensemble of sufficiently long integrations one can reduce 
initial-condition variability, and by keeping external factors such 
as greenhouse gas emmisions annually-periodic one can reduce 
variability due to non-stationarity.
A choice of metric with strong properties (e.g.\ (1)-(4)) is then 
crucial for interpretation of the
resulting bounds on rounding error in order to have confidence in the
reliability of a low-precision model. 
Although we developed our methods to measure 
rounding error, we hope they might also be of interest to
the broader climate modelling community.

\section{The Lorenz system}
\label{sec:L63}


\begin{figure*}[t]
\centering
	\includegraphics[width=.8\linewidth]{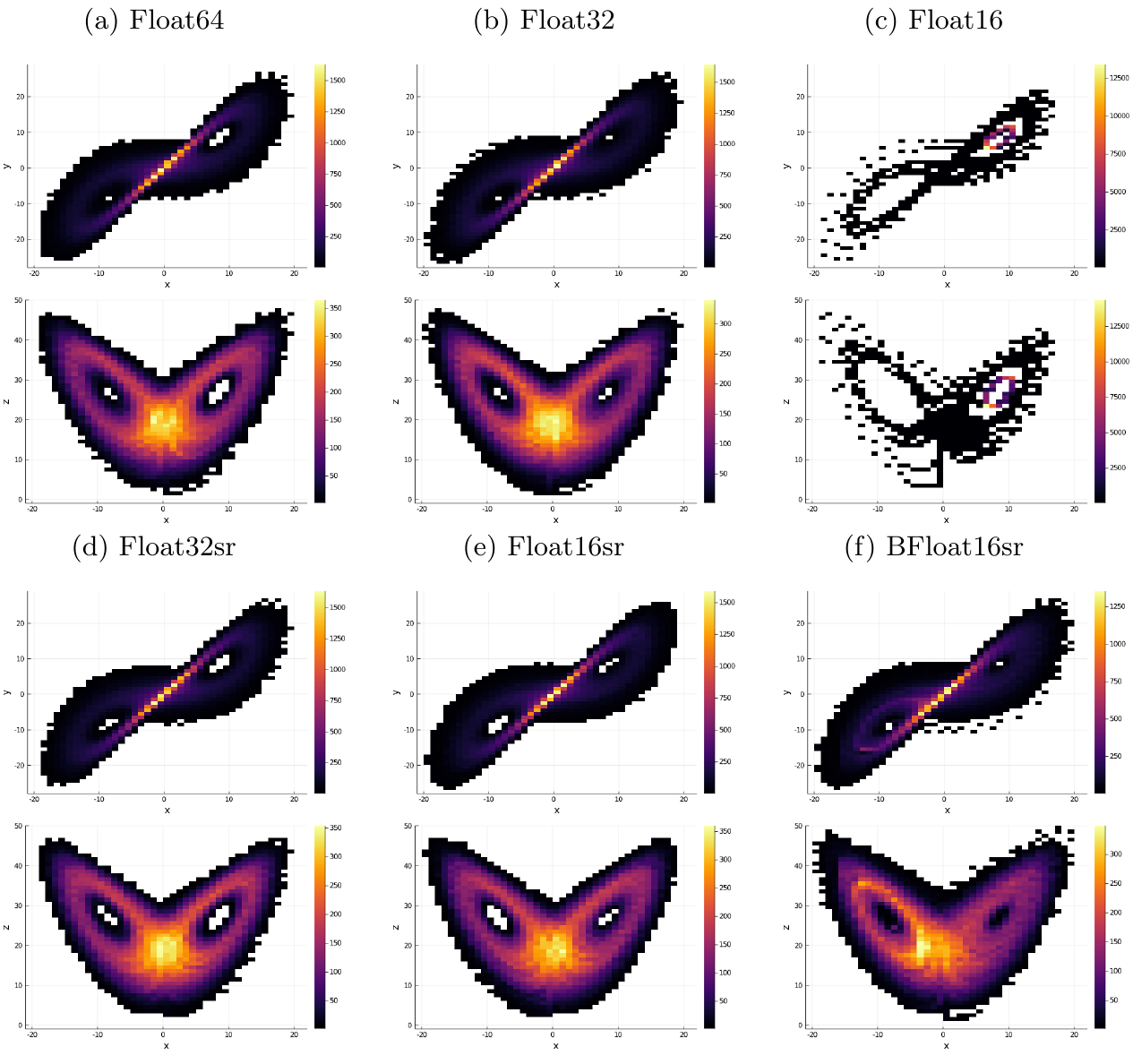}
	\captionsetup{width=0.9\linewidth}
	\caption{The Lorenz system integrated and data-binned at different precisions}
	\label{fig:L63histos}
\end{figure*}


To illustrate our methods which will 
later be applied to more complex climate models, we first consider the Lorenz system
\eqq{\label{L63} \frac{dx}{dt} &= 10\left(y - x\right) \\
\frac{dy}{dt}  &= \big(\nicefrac{8}{3} - z\big)x - y \\
\frac{dz}{dt}  &= xy - 28 z.} 

Derived first by Lorenz \cite{L63} in a study of convection, 
\eqref{L63} exhibits features of nonlinear dynamics
representative of the real atmosphere \cite{Molteni2018}. 
The system has a fractal attractor $\AAA\subseteq \RR^3$ and the dynamics on $\AAA$ is chaotic, rendering the precise 
numerical approximation of any specific orbit futile. On the other hand, a study of 
asymptotics on $\AAA$ reveals statistics common across orbits 
which one may hope to approximate. 
Indeed, there exists a unique invariant probability 
distribution $\nu$ supported on $\AAA$ such that for almost any orbit $\mathbf{x}(t)=(x(t), y(t), z(t))$ initiated in 
the basin of attraction of $\AAA$ and any bounded continuous $\phi$
\eq{\label{invariant_measure_identity} 
\lim_{T\uparrow \infty} \left\{ \frac{1}{T} \int_0^T \phi(\mathbf{x}(t)) dt \right\} = \iiint \phi(\mathbf{x}) d\nu(\mathbf{x})}
cf.\ \cite{TuckerExistence}. Thus $\nu$ encodes the long-time statistics of the system. 
For example, taking $\phi(\mathbf{x})=1$ for $\mathbf{x}\in B$ and $\phi(\mathbf{x})=0$ outside of a neighbourhood of
$B$, from \eqref{invariant_measure_identity} we see the average 
time an orbit spends in a region $B$ is the probability mass $\nu(B)$. 
In the context of climate modelling, the test for a 
low-precision integration of \eqref{L63} is whether it 
produces approximately the same 
$\nu$ as its high-precision counterpart.



We first sampled 10 initial conditions $i_0, \ldots, i_9\in \RR^3$
from a normal distribution with unit variance centred at the centre of mass of 
the attractor $\AAA$. We then integrated \eqref{L63} at high-precision ({Float64}) 
initialised at initial conditions $i_0,\ldots, i_4$ for $220\hskip2pt 000$ model time 
units (mtu) each and discarded the first $20\hskip2pt 000$\,mtu as spin-up to 
allow for any orbits initially perturbed off $\AAA$ to return to $\AAA$, and we labelled 
these five runs as the control ensemble 
$e^{\mathrm{control}}_i$. 
Next, for each comparison arithmetic---{including} {Float64}---we integrated \eqref{L63} initialised at initial conditions 
$i_5, \ldots, i_9$
for $220\hskip2pt 000$ mtu, discarded the first $20\hskip2pt 000$ mtu and 
labelled these as the competitor $e_j$. 
Each integration used the Runge-Kutta 4th order scheme with a time 
step of $dt=0.002$. 
For background on the different arithmetic formats and stochastic rounding, 
see App.\ \ref{app:Numbers}. 

In general, results will be sensitive both to the choice of numerical scheme 
and the time-step, however we won't dwell on such issues since our 
aim is to develop a method to measure rounding error in a climatological context, rather than to obtain 
an optimal integration.


\begin{figure}
	\includegraphics[width=\linewidth]{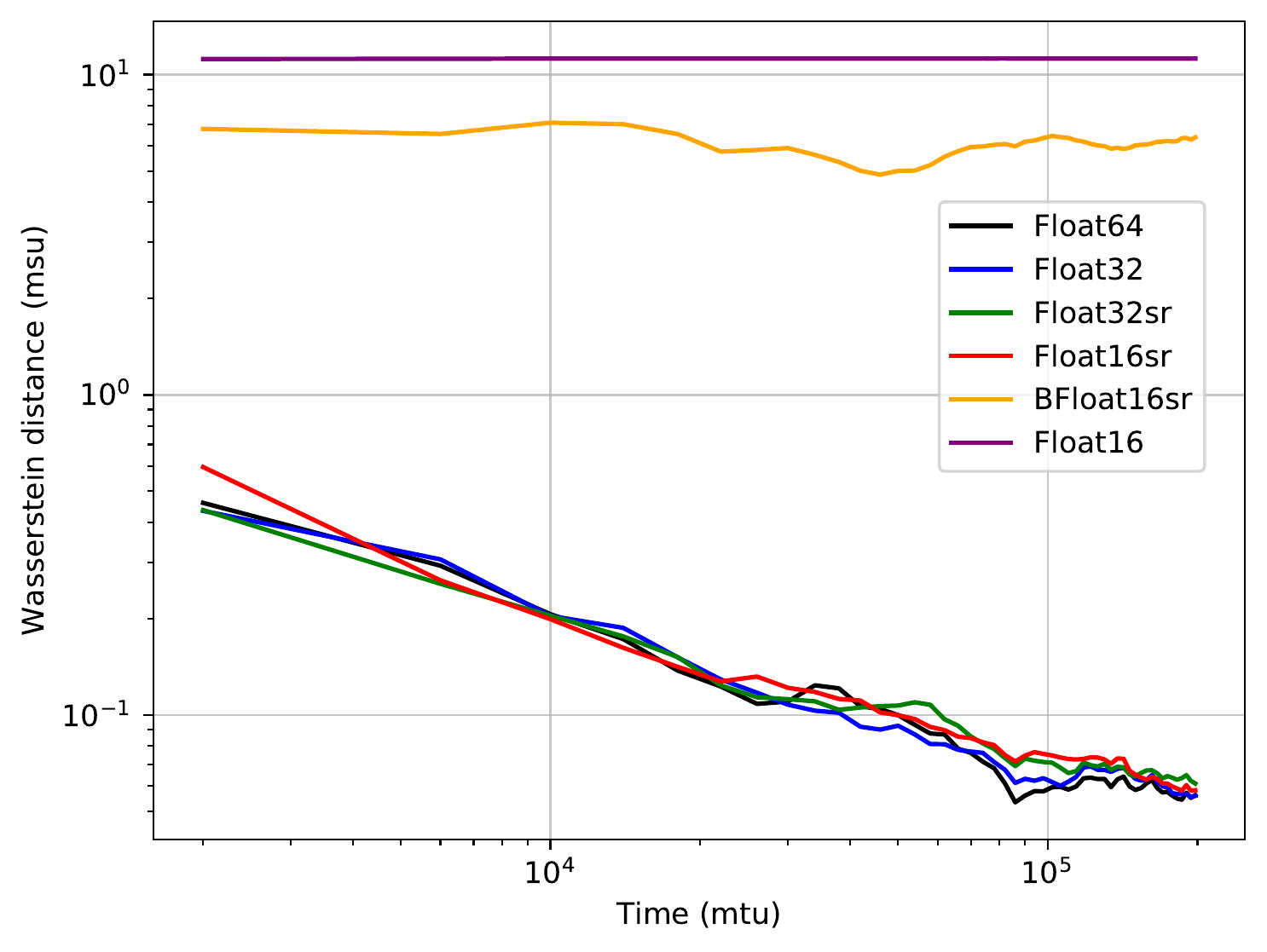}
	\captionsetup{width=0.9\linewidth}
	\caption{Measuring rounding error in the climate of the Lorenz system. 
	Plot shows 
	Wasserstein distances between the probability distributions generated by
	low-precision and high-precision ensembles (binwidth 6.0) where 
	black is high precision vs high precision for reference.}
	\label{fig:mainL63results}
\end{figure}


Integrations are binned and plotted 
in Fig.\ \ref{fig:L63histos}. 
Whilst the {Float32}, {Float32sr}, and {Float16sr} integrations appear to approximate 
the high-precision attractor well,
{Float16} suffers from the small number 
of available states forcing the evolution into an early periodic orbit 
while we found {BFloat16} collapsed on a point attractor with this fine time-step, 
though both of these integrations were notably improved by stochastic rounding. 
For each arithmetic we computed the ensemble mean 
Wasserstein distance (WD, cf.\ App.\ \ref{app:Wass}) between the 5 probability distributions 
generated by $e_i$ and the 5 
distributions generated by
$e^{\mathrm{control}}_j$ and
the evolution of this quantity with time is plotted in Fig.\ \ref{fig:mainL63results} 
with a log-log scale.
Note that the {Float64} competitor (black in Fig.\ \ref{fig:mainL63results}) 
shows the mean WD between a pair of high precision integrations
initiated at different initial conditions, and thus
gives a measure of the variability of the experiment 
at high precision which is 
important to be able to draw conclusions. 
Fig.\ \ref{fig:mainL63results} confirms quantitatively what is 
suggested by Fig.\ \ref{fig:L63histos} 
in that the {Float32}, {Float32sr} 
and {Float16sr} curves 
closely follow {Float64} in the approach to statistical equilibrium, 
showing that rounding error is small relative to high-precision variability, 
whilst for BFloat16sr rounding error has notably perturbed the dynamics.
The size of the high-precision variability after 100 000\,mtu 
(cf.\ Fig.\ \ref{fig:mainL63results}) is less than 0.1 model space 
units (msu) which is small
in the context of distributions supported on the Lorenz attractor, which 
has characteristic length scale of approximately 50\,msu 
(cf. Fig.\ \ref{fig:L63histos}). 


\begin{figure*}[t]
\centering
	\begin{subfigure}{.5\linewidth}
		\centering
		\includegraphics[width=\linewidth]{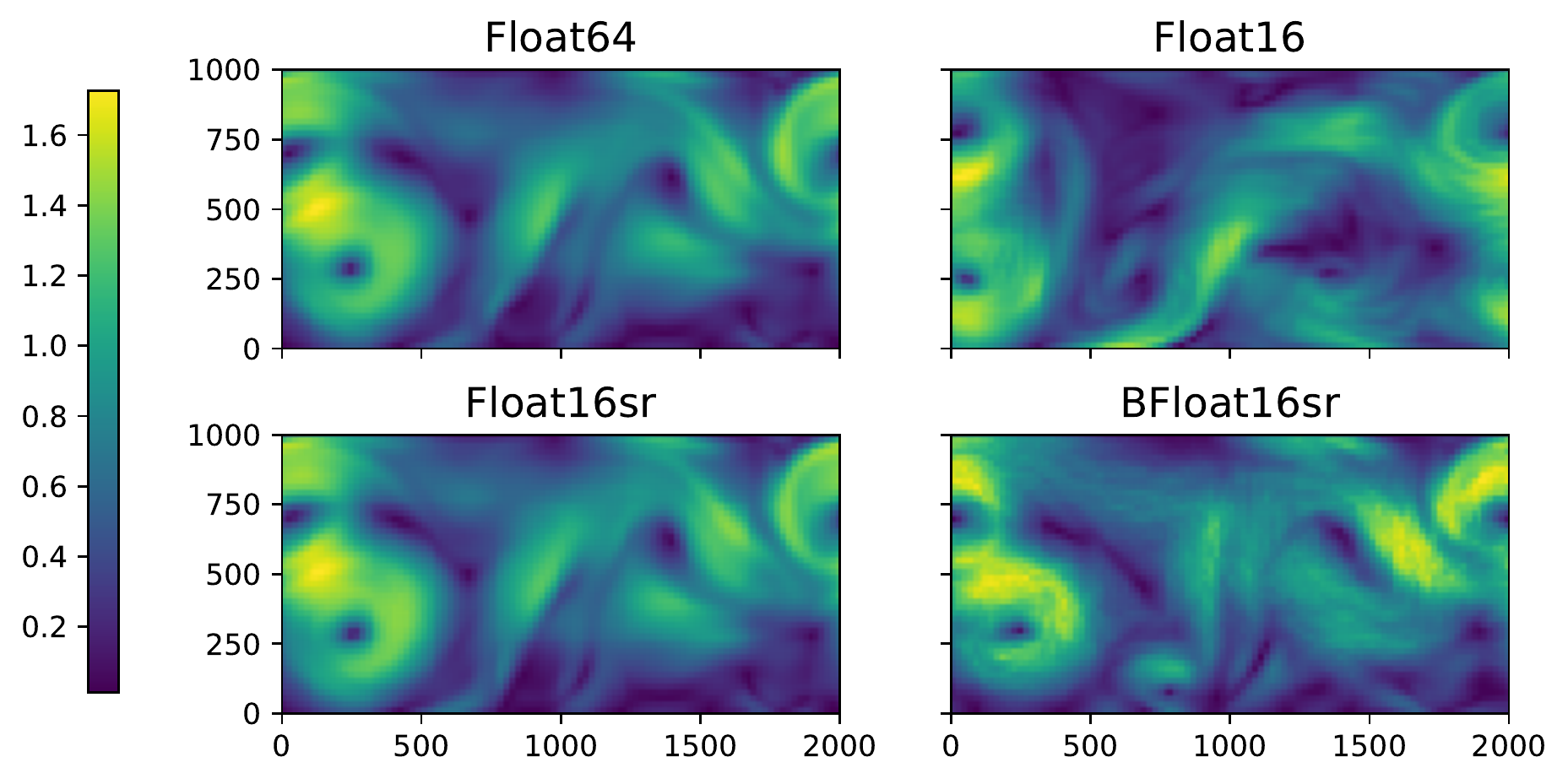}
		\captionsetup{width=.9\linewidth}
	\end{subfigure}
	\begin{subfigure}{.45\linewidth}
		\centering
		\includegraphics[width=\linewidth]{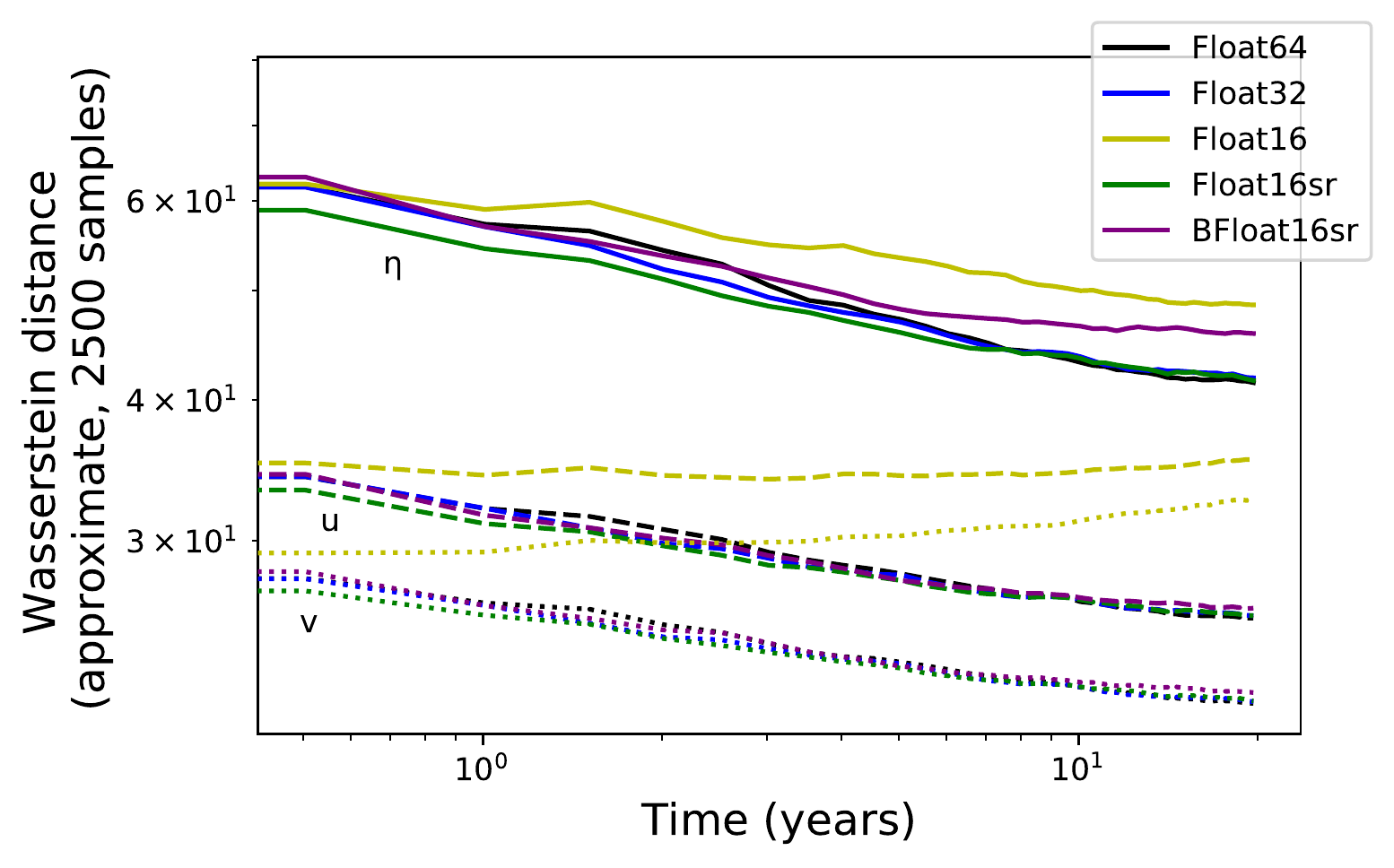}
		\captionsetup{width=.9\linewidth}
	\end{subfigure}
\captionsetup{width=.8\linewidth}
\caption{The shallow water model integrated at different precision levels. 
	The left-hand plot is a snapshot of the flow speed (m/s), initiated from the 
	same initial condition, after 50 days (a deterministic forecast) 
	while the right-hand graph 
	shows the effect of rounding error on climatology. 
	Double precision variabilty in black provides a reference
	and errors are small relative to this variability for 
	similarly converging distances.}
\label{fig:SWM}
\end{figure*}


To compute WDs we approximated the probability distributions by data-binning 
with cubed bins and a binwidth of 6\,msu. 
This is a coarse estimation but we found results were not sensitive to decreasing bin-width 
(in agreement with \cite{Vissio2018}) and we
also performed the same computation approximating by 
the empirical distributions generated by 2500 samples
as well as approximating by Sinkhorn divergences, and by marginalising onto 1-dimensional distributions (cf.\ \ref{app:exact_computation}), all of
which gave analogous results.

\section{A shallow water model}
\label{sec:SWM}

We next consider the shallow water model from \cite{Kloewer2020JAMES} 
which describes turbulent
flow in a rectangular ocean basin, driven by a steady zonally 
symmetric wind forcing over a meridionally symmetric ridge. The equations are
\al{
\frac{\partial \mathbf{u}}{\partial t} + (\mathbf{u} \cdot \nabla) \mathbf{u} +
f\hat{\mathbf{z}} \times \mathbf{u} &= -g\nabla \eta + \mathbf{D} + \mathbf{F} \\
\frac{\partial \eta}{\partial t} + \nabla \cdot (\mathbf{u}h) &= 0
}
where $\mathbf{u} = (u,v)$ is velocity, 
$\eta$ is surface elevation,
$\mathbf{D} = -\frac{c_D}{h}|\mathbf{u}| \mathbf{u} - \nu \nabla^4 \mathbf{u}$
is a nonlinear diffusive term with coefficient $\nu$ and $c_D$ a dimensionless parameter, 
$\mathbf{F}$ is wind forcing, $f$ is the Coriolis parameter,
$h=H+\eta$ is layer thickness 
and $H$ is the time-independent depth of the water at rest 
describing the ridge at the fluid base. The ocean basin dimensions were taken as 2000\,km by 1000\,km with average depth 500\,m.
We integrated the equations using the scheme from \cite{Kloewer2020JAMES} 
which uses finite differences on an Arakawa C-grid and 4th order Runge-Kutta in
time combined with a semi-implicit scheme for the dissipative terms, with a time step of 6\,hr, and refer to 
\cite{Kloewer2020JAMES} for more details on the numerical scheme.

Following the methodology developed in  
Sec.\ \ref{sec:L63} 
we integrated
for 20 years discarding the first year of each run as spin-up, 
taking a 5 member ensemble for each arithmetic.
Snapshots of the evolution in each case
are plotted in Fig.\ \ref{fig:SWM}a. 
We computed ensemble mean pairwise WDs between 
the distributions generated by high and low-precision ensembles 
and plot this quantity evolving with time in 
Fig.\ \ref{fig:SWM}. 
We found that for {Float16} and {BFloat16sr} 
rounding error is significant whilst for {Float32} and {Float16sr}
rounding error is small relative to high-precision variability.
In particular, our results show
that rounding errors at half-precision
are successfully mitigated by stochastic rounding 
in this climate experiment. 
Again, we refer to App.\ \ref{app:Numbers} for background on 
these different number formats.

The main difference between the methods of this section 
and Sec.\ \ref{sec:L63} is that we considered
here a dynamical system which is high dimensional, so that approximating probability 
distributions is non-trivial. 
For Fig.\ \ref{fig:SWM}b we approximated the 
invariant distributions by taking 2500 uniformly distributed 
random samples in time and computed WDs 
between the corresponding emprical distributions (cf.\ \ref{app:exact_computation}). 
This approximation method does not give readily interpretable results due
to a curse of dimensionality (\ref{app:exact_computation}) however we
obtained analagous results by marginalising onto one-dimensional subspaces. 
We save discussion of such
marginalised results for Sec.\ \ref{sec:SPEEDY} in the context of a 
global atmospheric model.

\section{Interlude: heat diffusion in an ideal soil column.}
\label{sec:heat}

In this section we briefly consider a very simplified land-surface 
component of a global climate model. 
This is a trivial case of 
climatology since all solutions converge upon a constant equilibrium temperature 
and so there is no need to use the WD in this setting. 
We include this simple example, however, because it clearly illustrates a major 
advantage of SR over RN---preventing stagnation.

This section was partially motivated by \cite{canadianMO} 
and \cite{Dawson2018}. 
In \cite{canadianMO} the authors observed that the Canadian LAnd 
Surface Scheme (CLASS) could not be run effectively at single-precision
in large part because of an issue of stagnation. 
They argued that single-precision arithmetic was not appropriate for climate 
modelling with the scheme which relies crucially on accurate representations 
of slowly-varying processes such as permafrost thawing and that 
double precision or even quadruple precision should be adopted instead. 
The set-up considered here was introduced in \cite{Dawson2018} 
as a toy model which retained some features of CLASS, 
most crucially the stagnation at single-precision with RN. 
The authors of \cite{Dawson2018} proposed mixed precision to avoid stagnation, 
while the results of this section indicate that SR 
provides an alternative approach. 

Following \cite{Dawson2018}, we consider an idealised 
soil column which is heated from the top and 
thermally insulated from the bottom
$$\frac{\dd T}{\dd t}= D \frac{\dd^2 T}{\dd z^2}$$
with $T(t,0) = 280$, $\frac{\dd T}{\dd z}(t,H) = 0$ and $T(0,z) = 273$
where $T(t,z)$ is temperature in Kelvin, $H=60$\,m is soil depth, 
and $D$=$7\times10^{-7} \mathrm{m}^2\mathrm{s}^{-1}$ is the coefficient of diffusivity, 
and discretise as 
\eq{
\label{bad_scheme}
T^{n+1}_j = T^n_j + D \Delta t \left( \frac{T^n_{j+1} -2T^n_j + T^n_{j-1}}{(\Delta z)^2} \right)
}
with $\Delta z = 1$\,m and $\Delta t = 1800$\,s.

\begin{figure}
\centering
\includegraphics[width=.95\linewidth]{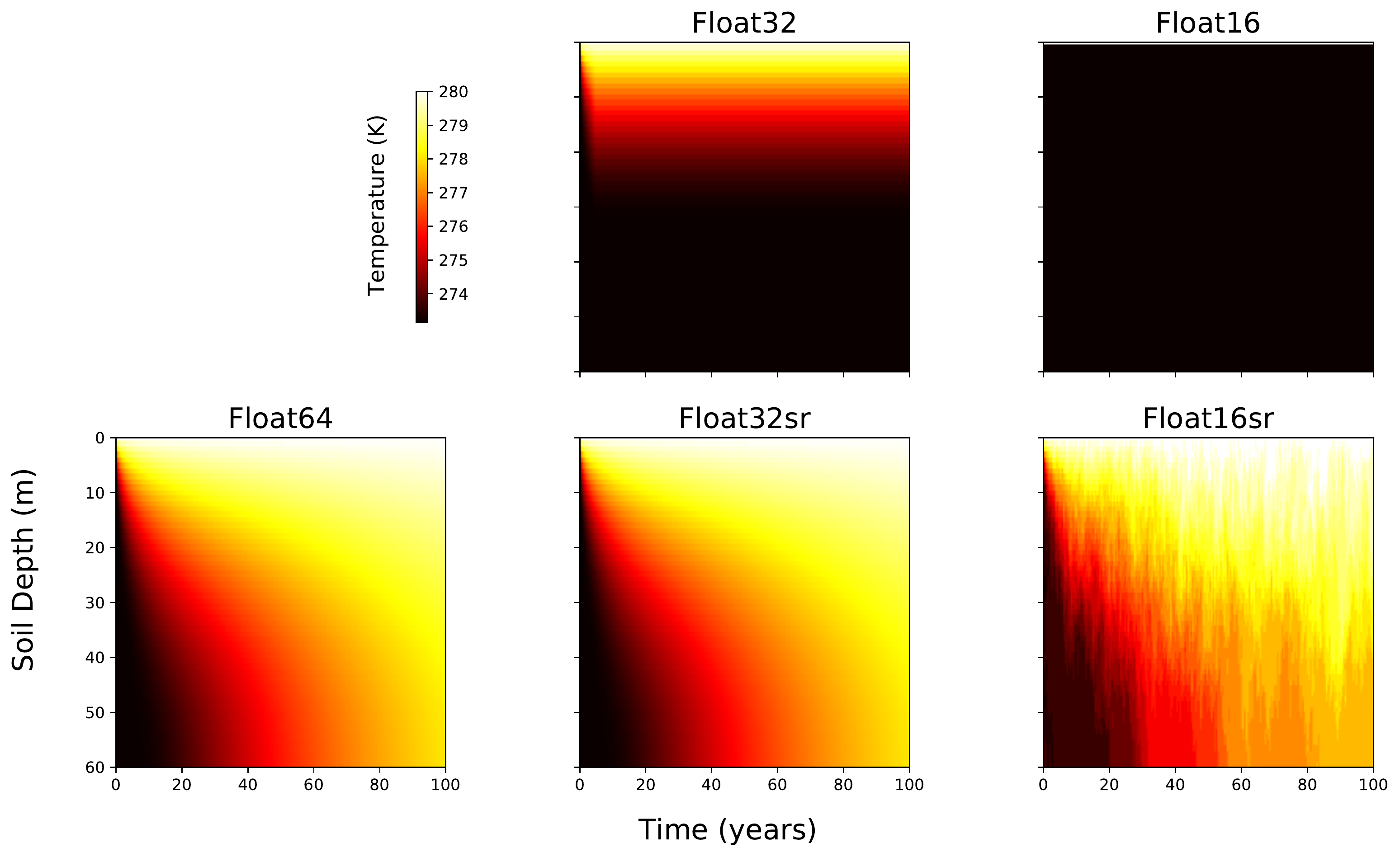}
\captionsetup{width=.9\linewidth}
\caption{Heat diffusion in a soil column with different number formats and rounding modes.}
\label{fig:heateqSR}
\end{figure}

We integrated for 100 years and 
the results are plotted in Fig.\ \ref{fig:heateqSR}. 
Stagnation is apparent for {Float32} and {Float16} 
where the small tendancy term in \eqref{bad_scheme} is repeatedly rounded 
down to zero, so that 
heat does not diffuse effectively through the soil column.
This is mitigated by SR, however, which assignes a non-zero 
probability of rounding up after the addition in \eqref{bad_scheme} (cf.\ Sec.\ \ref{sec:rounding}). 
Rounding error is neglegible with {Float32sr} and while 
visible as noise in {Float16sr}, the 
solution shares the large-scale pattern of {Float64}. 


To be clear, this section is not intended to imply that SR is necessary 
in the low-precision integration of the heat equation. 
There are other ways to avoid stagnation, such as increasing 
the time-step which is extremely small in this example and 
well below what is necessary for stability, 
or implementing a compensated summation for the timestepping. 
Rather, this section aims to illustrate an interesting advantage of SR 
in mitigating stagnation, by means of a clear and visual example.
For more analysis of SR in the numerical solution of the heat equation see \cite{croci2020}.



\begin{figure*}[t]
\centering
	\begin{subfigure}{\textwidth}
		\includegraphics[width=\textwidth]{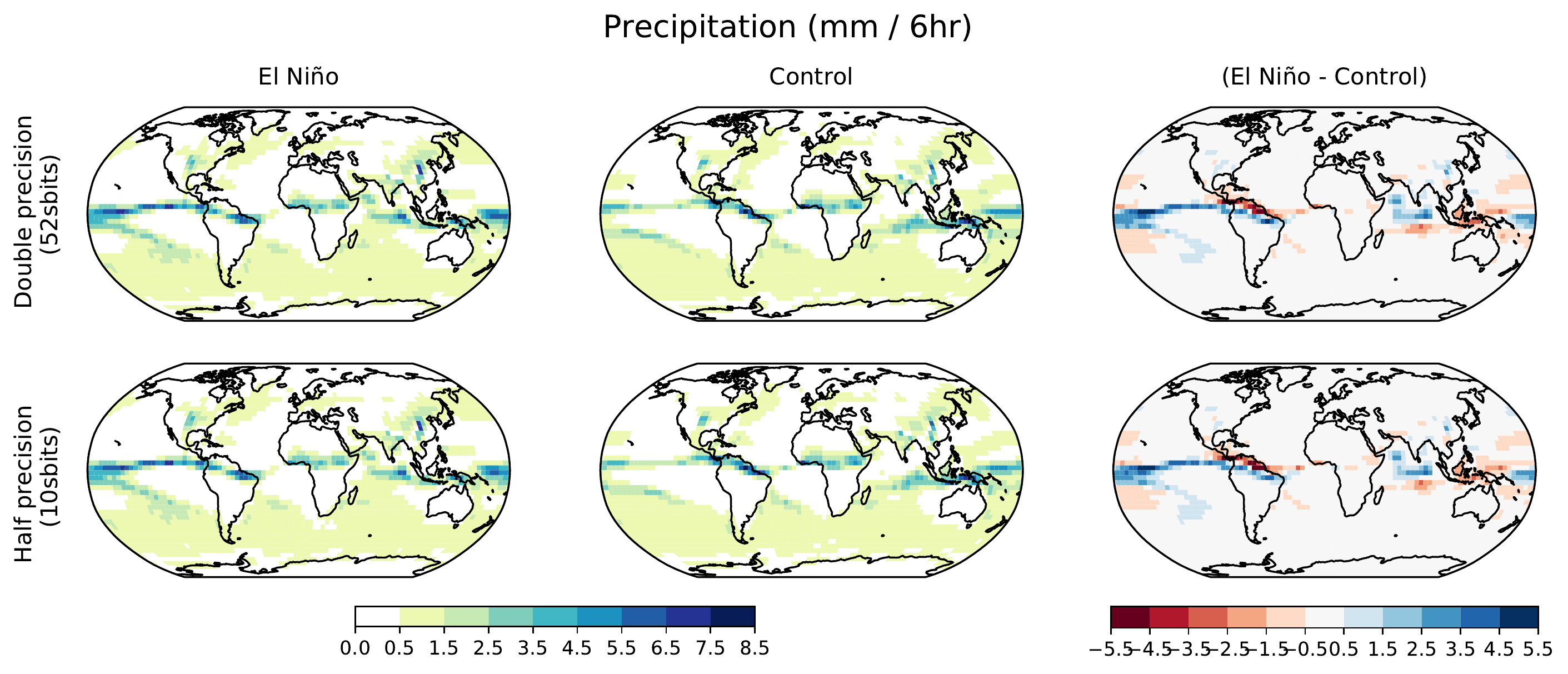}
	\end{subfigure}
	\\
	\begin{subfigure}{\textwidth}
		\includegraphics[width=\textwidth]{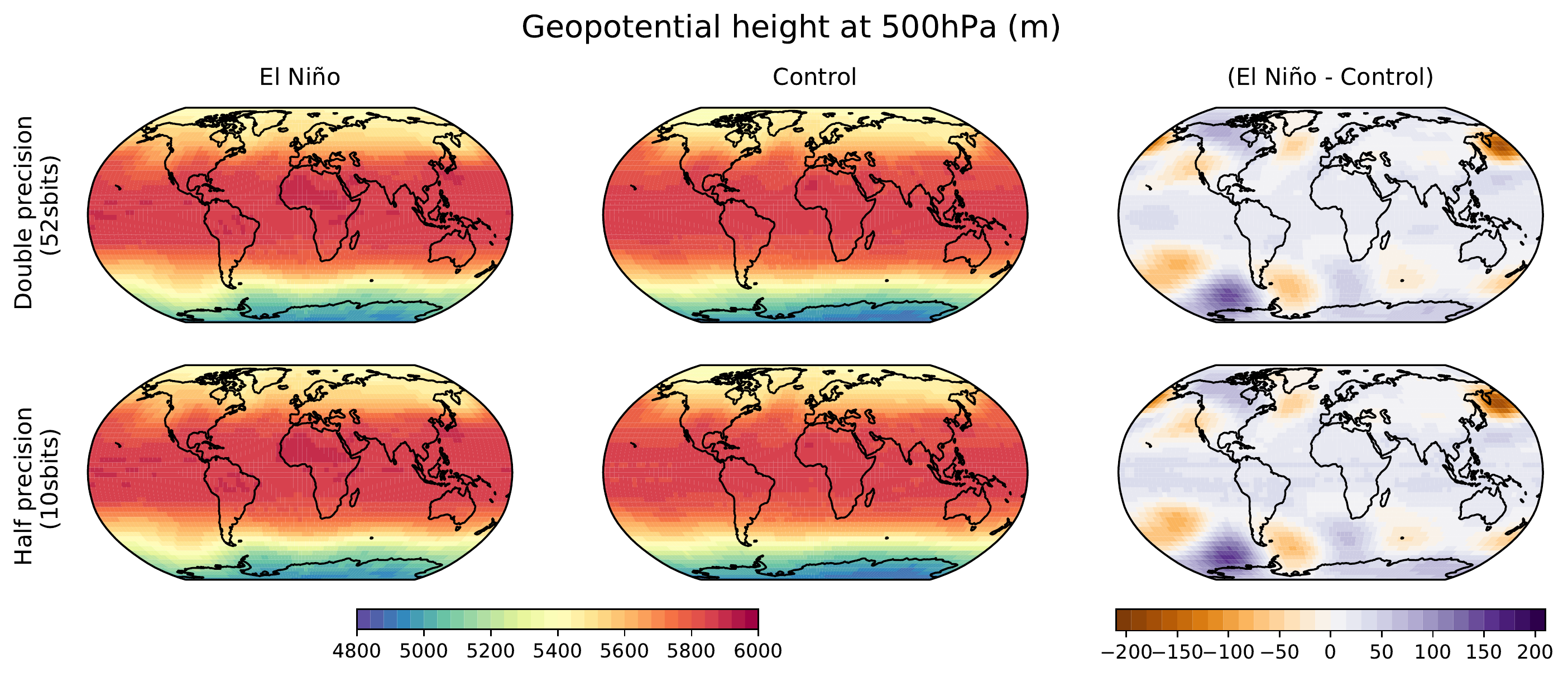}
	\end{subfigure}
\captionsetup{width=0.9\linewidth}
\caption{Annual mean response to El Ni\~no: double vs half precision. 
From the perspective of the mean state, rounding error 
at half precision is seen to be small.}
\label{fig:mean_states}
\end{figure*}


\section{A global atmospheric model.}
\label{sec:SPEEDY}
Finally, we proceed to a global atmospheric circulation model: 
the Simplified Parameterizations PrimitivE 
Equation DYnamics ver 41 (SPEEDY). 
SPEEDY is a coarse resolution model employing a T30 spectral truncation 
with a 48x96 latitude-longitude grid, 
8 vertical levels and a 40\,min timetep, and is 
forced by annually-periodic fields obtained from ERA reanalysis together 
with a prescribed sea surface temperature anomaly \cite{Kucharski2013}. 
For this section, in order to isolate the effects of numerical precision 
we truncated only the significant bits, 
so that when we speak of half precision, for example, we refer to 10 significant 
bits (sbits) and 11 exponent bits rather than the IEEE754 5 exponent bits (i.e.\ overflows were not accounted for in this section). 


\begin{figure*}[t]
	\centering
	\includegraphics[width=\textwidth]{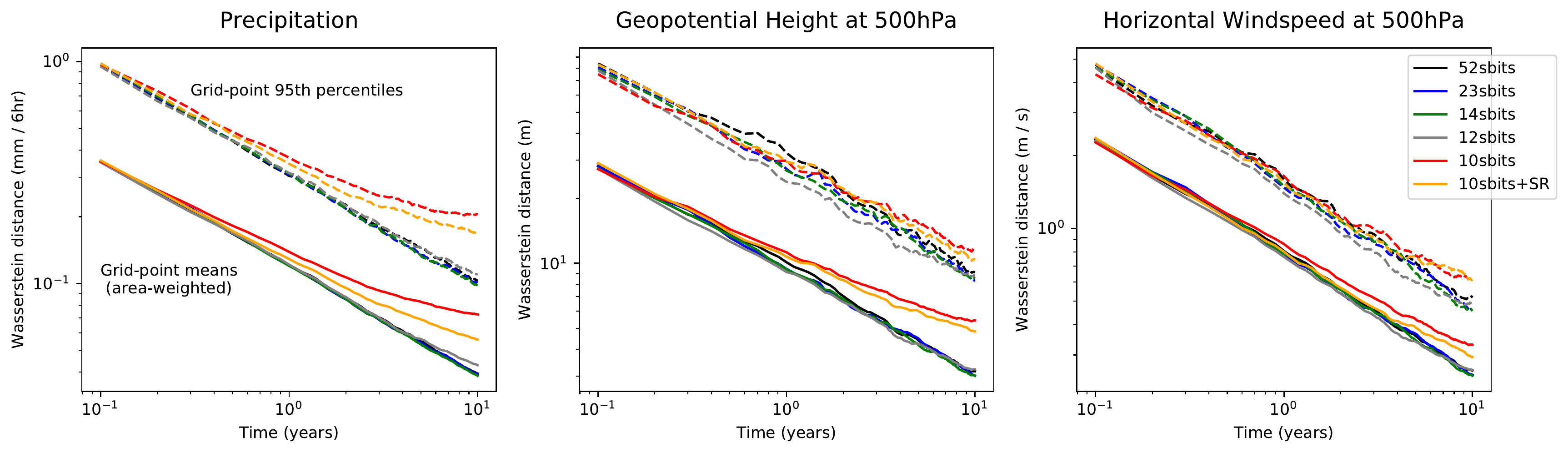}
	\captionsetup{width=.8\textwidth}
	\caption{Measuring rounding error in SPEEDY climatology with 
	grid point Wasserstein distances. 
	Double precision variabilty in black provides a reference
	and errors are small relative to this for 
	similarly converging distances.}
	\label{fig:GPWDconvergence}
\end{figure*}


As a first test, we constructed a constant-in-time SST anomaly field 
to crudely simulate an El Ni\~no event 
and ran SPEEDY both with and without to investigate the mean field response. 
This anomaly field was constructed (partially following \cite{Dogaretal})
by taking the Pearson correlation coefficients between an 
ERA reanalysis time series at each grid point and the Ni\~no 3.4 index 
over 1979-2019 and multiplying by a factor of 4 in an attempt to 
produce temperature anomalies in Kelvin roughly of the magnitude of the 2015 El Ni\~no. 
Fig.\ \ref{fig:mean_states} shows the El Ni\~no response for precipitation 
and geopotential height at 500\,hPa (Z500) for both double and half precision
and it is seen that the latter certainly simulates the response well. 
The area-weighted Pearson correlations between the 
double precision and half precision mean El-Ni\~no responses for the Northern Exatropics, Southern
Exatropics, and Tropics were calculated as: $(0.99, 0.99, 0.99)$ 
for precipitation and $(0.98, 0.99, 0.93)$ for Z500.

To explore the full climatology, we next followed the WD calculations of Sections \ref{sec:L63} and \ref{sec:SWM}. 
We generated initial conditions $i_0, \ldots, i_9$ by integrating from rest for 11 years at 51\,sbits of precision before
discarding the first year as spin-up and taking the 
initial conditions from the starts of each of the 10 subsequent years. 
This method was intended 
to emulate sampling from the high-precision invariant distribution 
whilst avoiding overlap in the high-precision ensemble. 
We then constructed our control ensemble $e_i^\textnormal{control}$ and 
competitor ensembles $e_j$
by integrating for 10 years from the initial conditions $i_0,\ldots,i_4$ and $i_5,\ldots,i_9$ respectively. 
The SST anomaly was turned off so that boundary conditions 
were annually periodic.

To circumvent issues of dimensionality we first marginalised onto 
the distributions spanned by individual spatial grid points and 
measure error by WDs between these 
1D distributions. 
We call these grid-point Wasserstein distances (GPWDs) and 
note this is the approach adopted in \cite{Vissio2020}.
To address correlations between grid-points,
we then checked our GPWD results against approximate WDs 
between the full distributions 
which were obtained via a Monte Carlo sampling approach
as was done in Sec.\ \ref{sec:SWM}. 
While such results are harder to interpret quantitatively
(cf.\ \ref{app:exact_computation})
we found that they were analogous to the GPWD results.
In particular, no errors were detected by this method which 
were not present in the GPWDs.

\begin{figure}[h]
\centering
	\includegraphics[width=\linewidth]{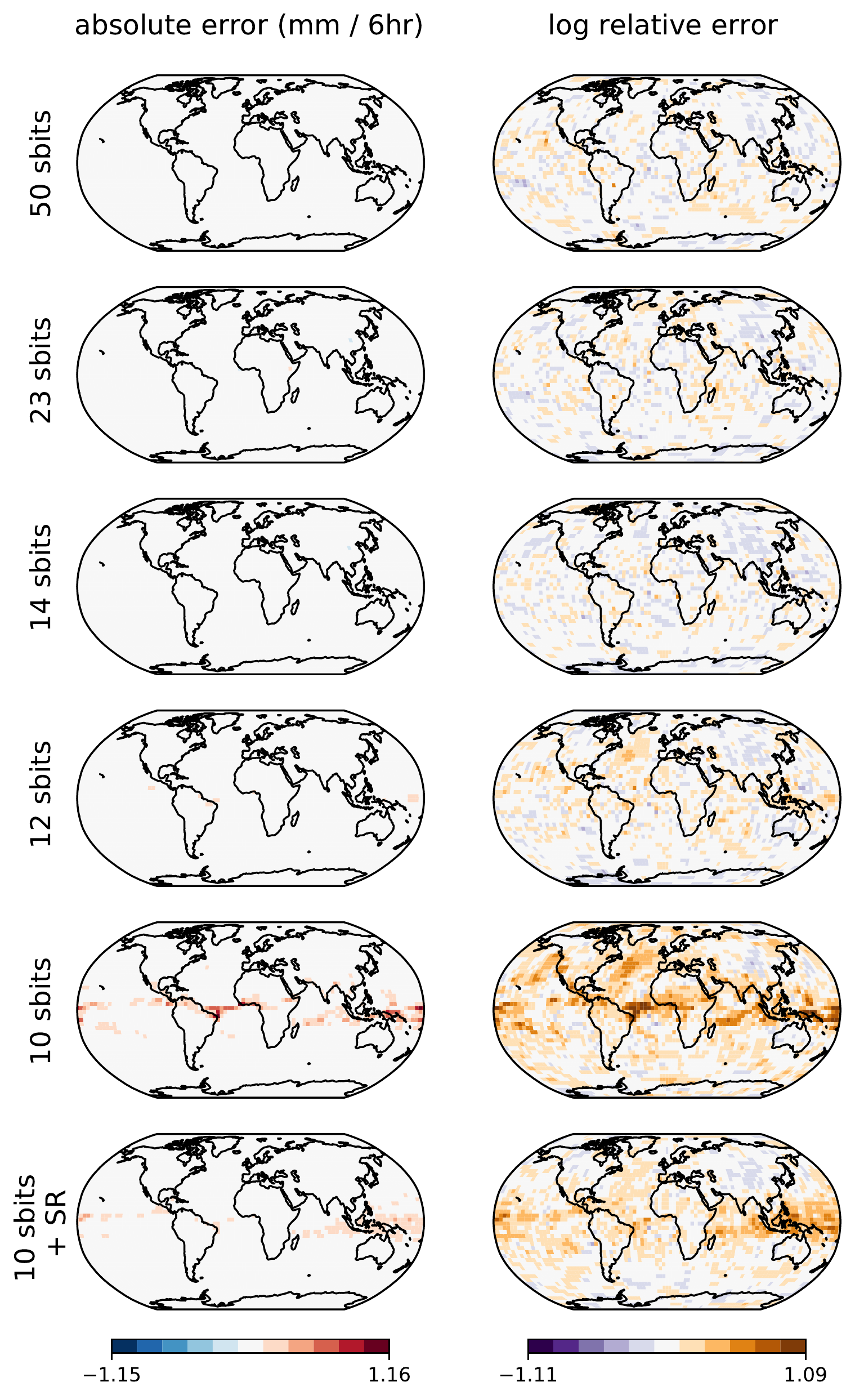}
	\captionsetup{width = .8\linewidth}
	\caption{Spatial distribution of precipitation error after 10 years (cf.\ \eqref{eq:AE}) \& \eqref{eq:LRE})}
	\label{fig:LRE_AE}
\end{figure}

The grid-point mean and 95th percentile GPWD results for total 
precipitation, Z500, and horizontal windspeed at 500\,hPa are plotted 
in Fig.\ \ref{fig:GPWDconvergence} as they evolve 
with time.
To give insight into the spatial distribution of rounding errors, 
Fig.\ \ref{fig:LRE_AE} shows maps of both the absolute error
\eqq{
\label{eq:AE}
\text{mean} \Big(& \WW(e^{\textnormal{low-precision}}_i, e^{\textnormal{control}}_j) \\
-& \WW(e^{\textnormal{high-precision}}_i, e^{\textnormal{control}}_j) \Big) }
and the log relative error
\eqq{
\label{eq:LRE} 
\log_{10} \left( 
\frac{ \text{mean}(\WW(e^{\textnormal{low-precision}}_i, e^{\textnormal{control}}_j)) }{ \text{mean}(\WW(e^{\textnormal{high-precision}}_i, e^{\textnormal{control}}_j) }  
\right)}
for precipitation (convective and large-scale combined) across
grid-points after the 10 year integrations have completed.


\begin{figure*}[t]
	\centering
	\begin{subfigure}{.49\linewidth}
		\includegraphics[width=\linewidth]{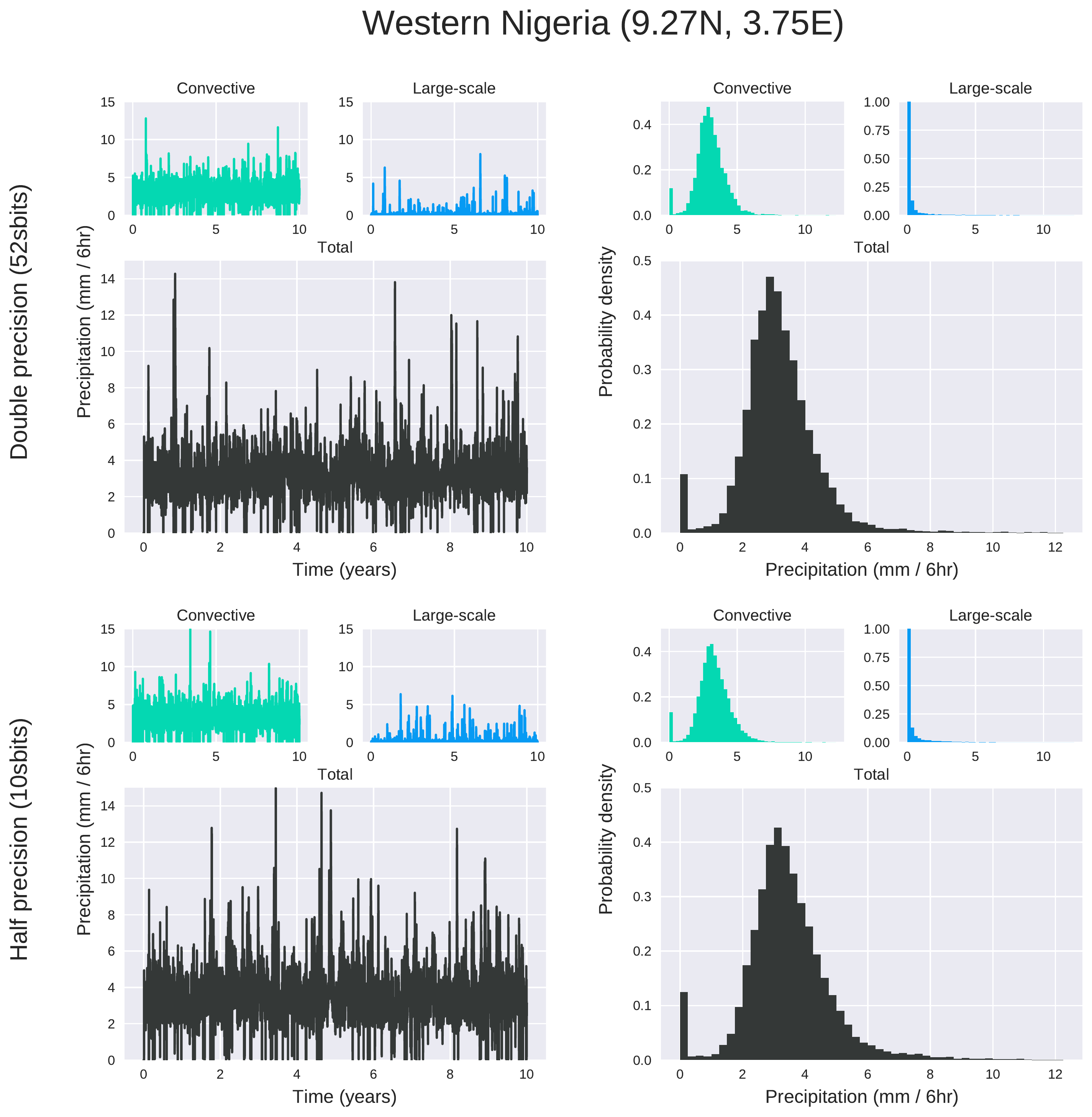}
		\label{fig:westnigeriarainfall}
	\end{subfigure}
	\begin{subfigure}{.49\linewidth}
		\includegraphics[width=\linewidth]{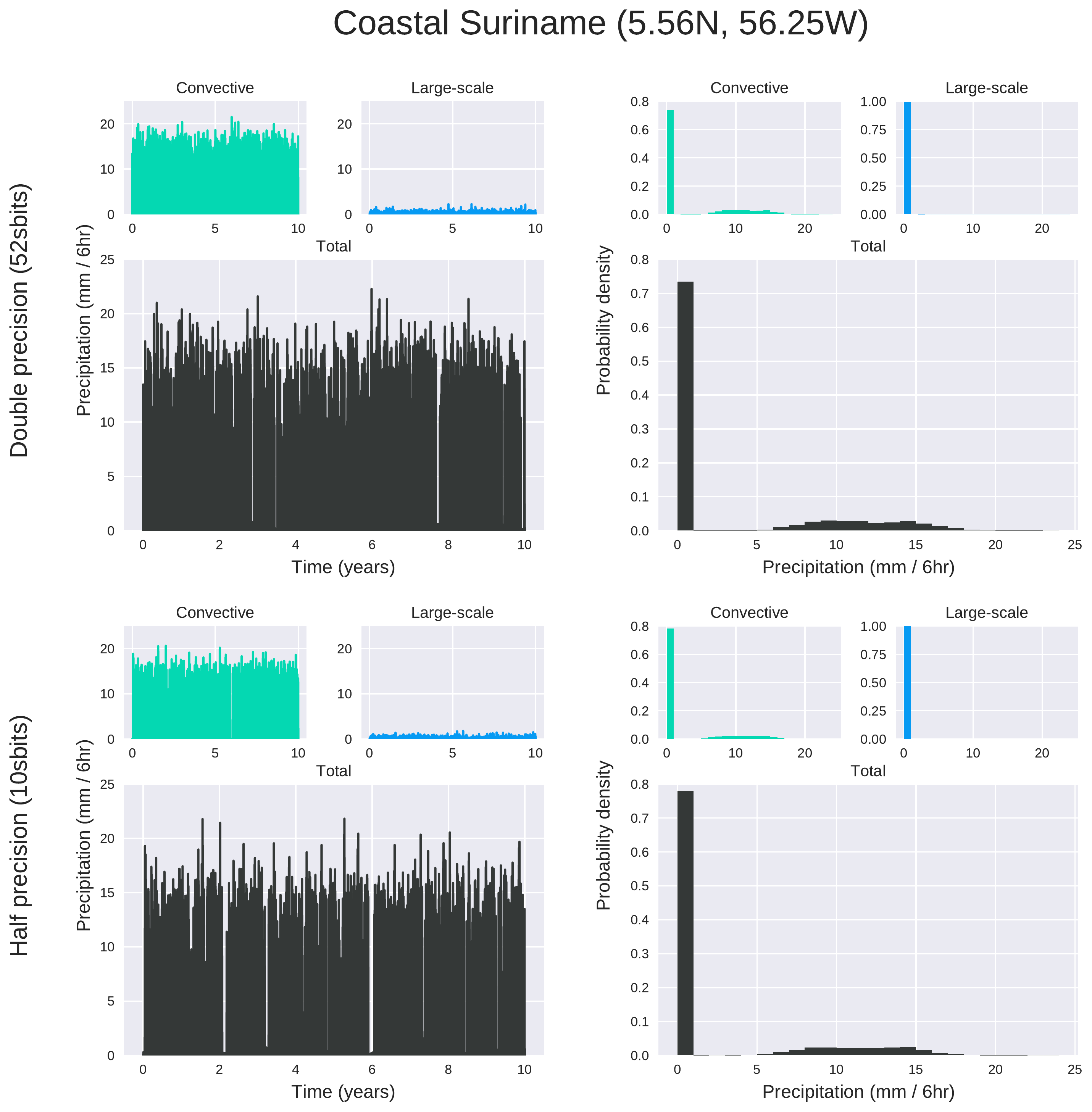}
		\label{fig:coastal_suriname}
	\end{subfigure}
	\captionsetup{width=.9\linewidth}
	\caption{Precipitation climatology at two grid points which are 
	representative of the largest rounding errors at half precision. 
	Even at these worst-affected grid points, the differences 
	between double and half precision are slight.}
	\label{fig:GPWD_case_studies}
\end{figure*}

For both geopotential height and horizontal 
windspeed we found that rounding error 
was neglegible relative to high-precision variability for 
12\,sbits and above, whilst a small rounding error emerged at 10\,sbits. 
For precipitation the picture was similar, except with a very small rounding error 
emerging at 12\,sbits.  
Fig.\ \ref{fig:LRE_AE} reveals those grid points at which 
rounding error becomes significant relative to high-precision 
variability for precipitation at 10\,sbits. 
Rounding error is neglegible relative to high-precision variability 
across all grid points for 14\,sbits and above, and
the mean high-precision variability for precipitation is around 
0.04\,mm/6hr which is very small 
(see \ref{sec:interpret_WD} for interpretation of WDs in terms of expected values).
Moreover, the rounding error at 10\,sbits is small, 
with grid-point mean values of 0.07\,mm/6hr, 5\,m and 0.3\,m/s for precipitation, 
geopotential height, and horizontal windspeed respectively, 
and with the worst affected grid points seeing errors of the order 
1\,mm/6hr, 25\,m and 1\,m/s respectively 
(recall that these values provide bounds on annual means 
as well as extreme weather events, cf. \ref{sec:interpret_WD}).
To give more intuition behind the size of rounding error at 10\,sbits, 
the probability distributions for precipitation at some of the worst 
affected tropical grid points (Coastal Suriname 5.56\,N, 56.25\,W and Western Nigeria 9.27\,N, 3.75\,E) are plotted in 
Fig.\ \ref{fig:GPWD_case_studies}. 
It is clearly seen that the difference between double and half precision, even at these worst affected grid points, is slight.

It may also be noted from Figs \ref{fig:GPWDconvergence} \& \ref{fig:LRE_AE} that stochastic rounding partially mitigates rounding error at half-precision.


We also computed differences in annual mean precipitation 
and found that, by and large, these were of the same order as the GPWDs, 
indicating that precipitation error was largely accounted for by differences 
in the means. In general, however, WD bounds are much 
stronger than mean bounds (cf.\ Sec.\ \ref{sec:interpret_WD}) 
so we can be confident that our estimates give stringent bounds 
on rounding error.

\section{Conclusions}
\label{sec:conclusions}
Whilst there is now convincing evidence that 
low-precision arithmetic can be suitable for accurate 
numerical weather prediction, before this work
there hadn't been a detailed study of the effects 
of rounding error on climate simulations,
and we have set out to address this imbalance.

We have argued that an appropriate metric to measure rounding error 
in the context of a chaotic climate model is the Wasserstein distance (WD), 
an intuitive and non-parametric metric which provides 
bounds on a range of expected values including those relevant 
for extreme weather events. 
By constructing experiments minimising the
variability between probability distributions at high precision and 
comparing WDs against low precision
we have obtained stringent bounds on rounding
error, and we have found that
error is typically insignificant until 
truncating as low as half precision in our climate experiments. 

We cannot conclude from our results that a state-of-the-art earth system
model can be run with equally low precision, since such codes are hugely 
complex which can make low precision issues difficult to overcome. 
However, given that the unit round-off error scales exponentially with 
the number of sbits, it would appear that the current industry standard 
of double precision across all model components is likely
overkill. 
In terms of acceptable precision, our results for
SPEEDY are similar to those found in an
analysis of the initial-value problem, 
suggesting that a level of precision suitable on weather timescales 
could also be suitable for climate---something which is not obvious a-priori. 
In light of recent operational successes with single-precision weather
forecasting, this is certainly a promising result in the direction of potential 
single-precision climate modelling.

Regarding stochastic rounding, although not currently in hardware, 
interest from machine learning together 
with a number of recently released patents suggest that it might
become available soon \cite{croci2020}. 
Our analysis has shown that stochastic rounding can make 
models more resilient to rounding error, especially at low precision 
though theoretically at any precision level, which 
gives some precedent for the community to 
engage now with chip-makers 
to secure the best hardware for next generation models.


\section*{Code}

For the Lorenz system we made use of \url{github.com/milankl/Lorenz63.jl}, 
for the shallow water model
\url{github.com/milankl/ShallowWaters.jl} version 0.4 and
for SPEEDY we used a branch developed by Saffin for 
which some changes were made to 
optimize for low-precision, as documented in \cite{Saffin2020}. 
We emulated low-precision using available types in Julia, in particular 
making use of \url{github.com/milankl/StochasticRounding.jl} for stochastic rounding, 
and we used the reduced precision emulator of \cite{DawsonDuben2016} in Fortran,
for which we developed a custom branch with stochastic rounding.
To compute optimal transport distances in 1D we used 
scipy.stats while for higher 
dimensional computations including Monte Carlo methods 
we built a custom solver at \url{github.com/eapax/EarthMover.jl}. 
Additional code gladly provided upon request. 

\section*{Acknowledgements}

The first author would like to thank Lorenzo Pacchiardi, Stephen Jeffress, Sam Hatfield, Peter D\"uben and Peter Weston for interesting discussions, as well as Lucy Harris for a careful reading of the manuscript. 

E.\ A.\ Paxton, M.\ Kl\"ower and T.\ Palmer were supported 
by the European Research Council grant number 741112,
M.\ Kl\"ower was supported by the Natural Environmental Research Council
grant number NE/L002612/1,
M.\ Chantry and T.\ Palmer were supported by a grant from the Office
of Naval Research Global,
and T.\ Palmer holds a Royal Society Research Professorship. 


\appendix
\section{The Wasserstein distance}
\label{app:Wass}
The Wasserstein distance (WD) defines distance 
between probability distributions 
$\mu$ and $\nu$ as the lowest cost at which one can transport 
all probability mass from $\mu$ to $\nu$ with respect to a cost function 
$c(x,y)=|x-y|^p$ which sets the cost to transport unit mass from position 
$x$ to position $y$, where in our work we have taken $p=1$ so that the WD 
has the units of the underlying field.
In this appendix we will first motivate the WD as a tool for the analysis 
of climate data by listing some of its favourable properties, before
giving the formal definition of the WD, discussing methods for its computation,
comparing it with other common metrics, and highlighting its 
interpretability through a useful dual formulation.

\subsection{Properties of the WD}
\label{WDproperties}

As stated earlier, the WD is defined (at least informally) as the smallest cost
required to transport one probability distribution into another. 
Before giving the formal details of this definition, let us first motivate the 
WD by lisiting some of its favourable properties.

First, the WD is non-parametric and versatile. 
It does not require any specific structure of the distributions such as Gaussianity 
and it can be used to compare both singular and continuous distributions. This is 
an important point for climate modelling which presents a wide range of 
probability distributions. 
For example, the climatological distributions corresponding to South Asian rainfall or the subtropical 
jet stream latitude are multimodal, while for the Lorenz system the 
object of interest is a singular probability distribution supported on a fractal attractor.
The ability to consider singular distibutions 
is also useful since it accomodates working directly with the empirical
distributions corresponding to a sample of data, rather than first binning the data into 
a histogram, for example.

Second, the WD is intuitive. It may be 
interpreted as the minimum amount of work required to transport one distribution 
into the other, an idea which is readily conceptualised, and it takes the 
units of the underlying field. 
For example, for distributions of rainfall measured in mm/day, a WD of 1 can be thought of as 
a difference of 1mm/day. 
Moreover, this figure provides bounds on differences in mean rainfall as well as
differences in extreme rainfall, for example 
(cf.\ \ref{sec:interpret_WD}).

Third, the WD takes into account geometry. 
To illustrate this, consider three simple distributions with densities
\begin{align*}
	f(x) = 	\begin{cases} 1 \quad x\in[0,1] \\
			0 \quad \text{otherwise} 
			\end{cases}
	\\
	g_1(x) = 	\begin{cases} 1 \quad x\in[1,2] \\
			0 \quad \text{otherwise} 
			\end{cases}
	\\
	g_2(x) = 	\begin{cases} 1 \quad x\in[9,10] \\
			0 \quad \text{otherwise.} 
			\end{cases}
\end{align*}
Now if $f$ is taken as the true distribution and $g_1$ \& $g_2$ as 
approximations of $f$, then clearly $g_1$ gives the better approximation, 
due to its proximity to $f$. 
This is reflected with $\text{WD}(f,g_1) < \text{WD}(f,g_2)$. 
By contrast, considering instead the $L^p$ distances between densities, 
for example, would give 
$\|f-g_1\|_{L^p}=\|f-g_2\|_{L^p}=2$ for all $p\geq1$. 
which does not reflect the geometry,
and this is only intensified in higher dimensions as is illustrated in 
Fig.\ \ref{fig:smiley}. 
This is not just a shortcoming of the $L^p$ metric but is shared by 
measures such as the Kolmogorov-Smirnoff test or the Kullback-Liebler 
divergence\footnote{The KL divergence is particularly ill-suited to comparing probability distributions on $\RR^d$ and this example gives $\mathrm{KL}(f,g_1)=\mathrm{KL}(f,g_2)=\infty$}. 

More generally, the WD metrises the space of probability distributions 
with respect to weak convergence \cite{villani2003topics}, 
which means that closeness in the sense of 
the WD corresponds to closeness with respect to a natural topology. 

\subsection{Defining the WD}

There are two alternative formulations of optimal transport due 
to Monge \& Kantorovich and it's helpful to consider both when 
computing WDs.

For the Monge formulation, suppose we have
discrete probability distributions 
\eq{\mu = \frac{1}{N} \sum_{i=1}^N \delta_{x_i} , \quad \eta = \frac{1}{N} \sum_{i=1}^N \delta_{y_i} }
where $x_i,y_i\in \RR^d$ and we visualize each as a distribution of 
equal masses on $\RR^d$. 
The masses might be books on a bookshelf $\RR$ or shipping 
crates on a dockside $\RR^2$. 
We are tasked with transporting $\mu$ to $\eta$. 
The masses cannot be split so a transport strategy is 
a permutation $\sigma$ of $N$ objects. 
Introducing a cost function $c(x,y)$
defining the cost to move unit mass from position $x$ to position $y$, 
the cost of $\sigma$ is $\frac{1}{N} \sum_{i=1}^N c(x_i,y_{\sigma(i)})$ 
and the optimal cost is the cost of an optimal strategy 
$\min_{\sigma\in S_N}\left\{ \frac{1}{N} \sum_{i=1}^N c(x_i, y_{\sigma(i)})\right\}$ 
where $S_N$ is the set of permutations. 
The special case $c(x,y)=|x-y|$ defines (Monge's version of) the WD
\eq{\label{mongeWD}W_1(\mu, \eta) = \min_{\sigma\in S_N} \frac{1}{N} \sum_{i=1}^N |x_i - y_{\sigma(i)}|}

Kantorovich's formulation is a relaxation of Monge's in that 
masses are viewed as continuous rather than discrete (think piles of 
sand rather than shipping crates) so mass can be subdivided in infinitely many ways. 
Suppose now a pair of distributions
\eq{\label{kantorovich_form}\mu = \sum_{i=1}^{M_1} P_i \delta_{x_i} , \quad \eta = \sum_{j=1}^{M_2} Q_j \delta_{y_j} }
where $\sum_i P_i = \sum_j Q_j = 1$ and $P_i, Q_j \geq 0$ (i.e. the $P$s and $Q$s are probability vectors).  
In applications $\mu$ and $\nu$ may represent discrete probability histograms, where the points 
$x_i$ and $y_j$ are the mid-points of bins and $P_i$ and $Q_j$ are weights\footnote{Kantorovich's 
formulation extends to more 
general distributions but 
we consider \eqref{kantorovich_form} to simplify things.}.

A transport strategy is now defined as a non-negative valued matrix $\pi\in \RR^{M_1\times M_2}_{\geq 0}$ 
where $\pi_{ij}$ denotes the amount of mass transported from $x_i$ to $y_j$ and 
for conservation of mass we impose $\sum_j\pi_{ij} = P_i$, $\sum_i \pi_{ij} = Q_j$. 
Write $c_{ij}$ for the cost to move unit mass from $x_i$ to $x_j$ so the cost of a 
strategy is $\sum_{i,j}c_{ij} \pi_{ij}$ and the optimal cost is $\min_{\pi\in\Pi(P,Q)} \sum_{i,j} c_{ij} \pi_{ij}$ 
where 
\eqq{\label{polytope}\Pi(P,Q&) = \bigg\{ \pi\in \RR^{M_1\times M_2}_{\geq 0} \colon \\ &\sum_{j} \pi_{ij} = P_i, \quad \sum_{i} \pi_{ij} = Q_j \bigg\}}
is the set of possible transport strategies. 
Note that the space $\Pi(P,Q)$ is also the space of joint distributions with marginals $P$ and $Q$, and $\Pi(P,Q)$ is 
non-empty as can be seen by considering the independence distribution $\pi_{ij}=P_iQ_j$.  
The special case $c_{ij}=|x_i-y_j|$ defines (Kantorovich's version of) the WD
\eq{\label{kantorovichWD}
W_1(\mu, \nu) = \min_{\pi\in\Pi(P,Q)}\sum_{i,j} |x_i-y_j| \pi_{ij}}

If $M_1=M_2=N$ and $P_i=Q_j=\frac{1}{N}$ it may be shown there is 
a minimizing $\pi$ for \eqref{kantorovichWD} which is an optimal strategy in the sense of 
Monge so \eqref{mongeWD} and \eqref{kantorovichWD} are consistent \cite[p5-6]{villani2003topics}.

\subsection{Computing the WD}
\label{app:exact_computation}

Suppose one has samples $\{x_i\}_{i=1}^N$ and $\{y_i\}_{i=1}^N$ drawn from 
a pair of distributions $\mu$ and $\nu$ on $\RR^d$. 
Then one can either compute the Monge WD \eqref{mongeWD} between the empirical distributions
$\mu_n = \frac{1}{N}\sum_i \delta_{x_i}$ and $\eta_n = \frac{1}{N}\sum_i \delta_{y_i}$ directly or 
one can first perform a data-binning of the data into $M$ bins and compute the Kantorovich 
WD \eqref{kantorovichWD} between the resulting histograms. 
The complexity of the former scales with the sample size $N$ while the latter 
scales with the number of bins $M$. 

The computation of \eqref{mongeWD} is a special case of the {assignment problem} from economics 
which can be solved in $\mathcal{O}(N^3)$ by the well-known Hungarian algorithm. 
On the other hand, the Kantorovich formulation \eqref{kantorovichWD} is an example of a 
problem in {linear programming}. 
The set $\Pi(P,Q)$ is a convex polytope and as the cost function is linear 
it follows that the minimum must be attained on a vertex of this polytope. 
A minimizing vertex can be found, for example, via the famous simplex algorithm. 

When $d=1$ the WD can be computed easily as there is 
an explicit formula. Indeed, for two 1D distributions with cumulative distribution functions 
(CDFs) $F$ and $G$ respectively the 1-WD is (\cite[p75]{villani2003topics})
\eq{\label{explicit_form1d}W_1(F,G) = \int_{-\infty}^\infty | F(x)-G(x) | dx.}

When $d$ is large data-binning is infeasible 
and it is more natural to work with the empirical distributions $\mu_n$, $\nu_n$ directly 
(resorting to the Monge formulation) 
however there is a curse of dimensionality in this context. Indeed, one has
\al{\mathbb{E}[ |W_1(\mu_n, \nu_n)-W_1(\mu, \nu)| ] = \mathcal{O}(n^{-\frac{1}{d}})}
and this bound is sharp in general
which gives very slow convergence in high dimension in 
some cases \cite{dudley1969}.
Construction of a metric to rival the WD which does not suffer a curse 
of dimensionality is an open problem and active research. 

In our work we have found that, despite the curse of dimensionality, 
computating the WD between empirical distributions with a modest 
sample size is computationally feasible and 
provides a useful checksum, usually in agreement with
results obtained for example by marginalising on one-dimensional subspaces. 
We also note that interesting recent work has shown a regularised form of the WD 
called the Sinkhorn divergence (SD) \cite{cuturi2013} has improved sample complexity with a
dimension agnostic convergence rate of $\mathcal{O}(n^{-\frac{1}{2}})$ for appropriate 
regularising parameters \cite{genevay2019} and in our work we found that many WD 
computations could be corroborated by SDs provided a suitable choice of regularising
parameter was chosen to ensure convergence of the Sinkhorn algorithm. 

\subsection{Comparing other metrics}
\label{sec:other_metrics}

It is interesting to note the similarity between \eqref{explicit_form1d} 
and the Continuous Rank Probability Score often used for weather forecast skill
\eq{\label{crps}\mathrm{CRPS}(F,G) = \int_{-\infty}^\infty | F(x) - G(x) |^2 dx}
and with the Kolmogorov-Smirnoff test
\eq{\label{KS}\mathrm{KS}(F,G) = \sup_{x\in(\infty,\infty)} |F(x) - G(x)| }
for CDFs $F$ and $G$. For the WD with $p=2$ cost there is 
the explicit formula in 1D
\eq{\label{Wp1d}W_2(F,G) = \int_{-\infty}^\infty |F^{-1}(x) - G^{-1}(x)|^2 dx }
where $F^{-1}$ and $G^{-1}$ are generalized inverses \cite[p.75]{villani2003topics}. 
Note that \eqref{explicit_form1d}, \eqref{crps} and \eqref{Wp1d} take account of the 
geometry of $\RR$ (cf.\ \ref{WDproperties}) while \eqref{KS} does not. 

\subsection{Duality.}
\label{sec:interpret_WD}

Suppose we have a pair of distributions $\mu$ and $\nu$
representing, for example, rainfall at a fixed location in mm/6hr. 
How can we interpret a WD of, say, 1 between $\mu$ and $\nu$? 

Since cost is defined as $c(x,y)=|x-y|$ a nice property of the WD is that it inherits the 
units of rainfall so that we can interpret the difference in mm/6hr. 
Heuristically, this difference tells us that a cost of at least 1mm/6hr
must be spent to transport $\mu$ to $\nu$,
and this takes into account both mean and extreme rainfall.

Moreover, this distance gives bounds on a range of expected values by the 
Kantorovich-Rubenstein duality \cite[Thm 1.14]{villani2003topics} which states
\al{
W_1(\mu,\nu) = \sup_{f\in \mathrm{Lip}_1} \left| \EE[f(X_\mu)] - \EE[f(Y_\nu)] \right| }
where $X_\mu$ and $Y_\nu$ are random variables 
with laws $\mu$ and $\nu$ and $\mathrm{Lip}_1$ is the space 
of functions satisfying $|f(x)-f(y)|\leq|x-y|$ (the 1-Lipschitz functions).
Taking $f(x)=x$ and duality gives
\al{
\left| \EE[X_\mu] - \EE[Y_\nu] \right| \leq W_1(\mu,\nu)}
which shows that a WD of $1$mm/6hr 
implies a difference in expected rainfall of less than 1mm/6hr (note this bound is sharp when $\mu$ and $\nu$ 
are Dirac masses). 
But duality also gives bounds on expected extreme rainfall. 
To see this, suppose extreme rainfall is defined as any rainfall which falls 
in excess of $r_c$ mm/6hr where $r_c$ is some critical value. 
Then taking $f(x) = 0$ for $x<r_c$ and $f(x)=x-r_c$ for $x \geq r_c$ gives
\al{\left| \EE[f(X_\mu)] - \EE[f(Y_\nu)] \right| \leq W_1(\mu,\nu)}
which shows a difference in expected extreme rainfall of $\leq1$mm/6hr.

Understanding such heuristics is helpful in interpreting the 
bounds on rounding error derived in this paper.

\section{Number formats}
\label{app:Numbers}
\subsection{Floating-point arithmetic}

\begin{figure*}[t]
	\centering
	\includegraphics[width=.9\linewidth]{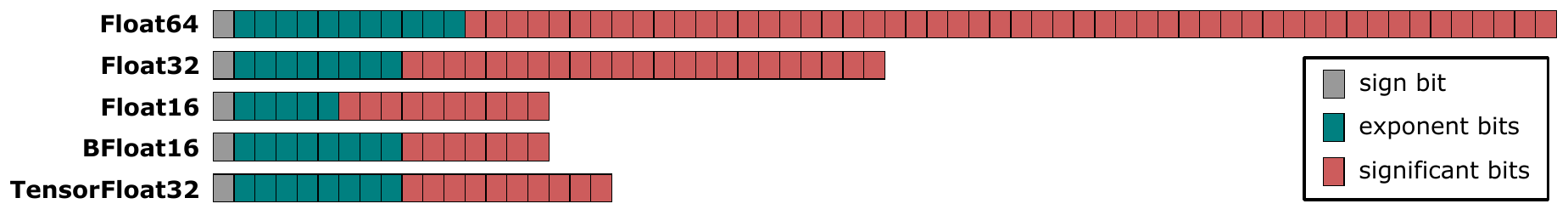}
	\captionsetup{width=.9\linewidth}
	\caption{Floating-point number formats available in hardware. 
	The IEEE754 standard for Float64 is 1 sign bit, 11 exponent bits 
	and 53 significant bits, whilst Float32 and Float16 break down 
	as (1, 8, 23) and (1, 5, 10) respectively. 
	Google's BFloat16 format (the B stands for Brain) opts for 
	a (1, 8, 7) split, sacrificing precision 
	for a larger dynamic range, while NVidia's new TensorFloat32 
	is (1, 8, 10) retaining the precision 
	of Float16 but with the dynamic range of Float32.}
	\label{fig:floats}
\end{figure*}

The standard arithmetic format for scientific computing
is the floating-point number (float).
The bits in a float are divided into three groups: 
a sign bit, the exponent bits, and the significant bits. 
A non-zero exponent specifies an interval $I=[2^{e-\text{bias}}, 2^{e+1-\text{bias}})$
with a bias to allow for negative exponents while the exponent bits
are interepreted as an unsigned integer $e$.
For $e=0$, floats are defined on an interval $I=(0, \pm2^{1-\text{bias}})$ called the subnormal range.
The signficant bits specify a point on $I$ from an evenly-spaced partition of $I$. 
Thus, for a given bias, the number of exponent bits determine
the dynamic range of representable numbers in the normal range, while
the subnormal range, and therefore the smallest representable number,
is determined by the number of significant bits.
Some different float formats available in hardware 
are shown in Fig.\ \ref{fig:floats}. 

The IEEE-754 Float64 format is called double precision, 
and Float32 and Float16 single and half precision respectively. 

\subsection{Rounding}
\label{sec:rounding}

The default rounding mode for floats is round-to-nearest tie-to-even (RN) which rounds an exact result 
$x$ to the nearest representable number $x_i$. In case $x$ is half-way between two representable numbers, 
the result will be tied to the even float, whose significand ends in a zero bit. These special cases are
therefore alternately round up or down, which removes a bias that would otherwise persist. 


For stochastic rounding (SR) rounding of $x$ down to a representable number $x_1$ or up to $x_2$
occurs at probabilities that are proportional to the respective distances. 
Specifically, if $u$ is
the distance between $x_1,x_2$, then $x$ will be rounded to $x_1$ with probability
$1 - u^{-1}(x - x_1)$ and to $x_2$ with probability $u^{-1}(x - x_1)$.

The introduced absolute rounding error for SR
is always at least as big as for RN
and when low-probability round away from nearest occurs, 
it can be up to $\pm u$, twice as large as for round-to-nearest. 
However, by construction, SR is exact in expectation 
and thus in particular by the law of large numbers one has
\al{\lim_{N\to \infty} \frac{1}{N} \sum_{i=1}^N \text{stochastic~round}(x) = x}
with the limit obtained in the strong sense. 
Moreover, by sometimes rounding small remainders up,
rather than always rounding them down as in RN, systemic 
errors can sometimes be avoided with SR, such as in
stagnation (see Sec.\ \ref{sec:heat} for an example).

It is worth noting that SR at low precision 
requires computation
at a higher precision in order to generate the probabilities for rounding, 
however all numbers are written, read, and communicated at 
low precision. 
It is also interesting to note that SR can easily be implemented with a random number 
sampled from the uniform distribution, which can thus be done in parallel to the arithmetic. 



%

\bibliographystyle{abbrv}
\bibliography{WDbibliography}

\end{document}